\documentclass[12pt]{article}
 \usepackage[paperheight=12in,paperwidth=8.75in]{geometry}

\usepackage{setspace}
\usepackage[T1]{fontenc}
\usepackage{times}
\usepackage{palatino}
\usepackage{lmodern}
\usepackage[latin1]{inputenc}
\usepackage{epsfig}
\usepackage[english]{babel}
\usepackage{color}
\usepackage{graphicx}
\usepackage{dcolumn}
\usepackage{moreverb}
\usepackage{amsmath,amssymb,amsfonts}
\usepackage[all]{xy}

\begin{document}
\numberwithin{equation}{section}
\newcommand{\boxedeqn}[1]{%
  \[\fbox{%
      \addtolength{\linewidth}{-2\fboxsep}%
      \addtolength{\linewidth}{-2\fboxrule}%
      \begin{minipage}{\linewidth}%
      \begin{equation}#1\end{equation}%
      \end{minipage}%
    }\]%
}


\newsavebox{\fmbox}
\newenvironment{fmpage}[1]
     {\begin{lrbox}{\fmbox}\begin{minipage}{#1}}
     {\end{minipage}\end{lrbox}\fbox{\usebox{\fmbox}}}

\raggedbottom
\onecolumn

\parindent 8pt
\parskip 10pt
\baselineskip 16pt
\noindent\title*{{\Large{\textbf{Algebraic calculations for spectrum of superintegrable system from exceptional orthogonal polynomials }}}}
\newline
\newline
Md Fazlul Hoque$^a$, Ian Marquette$^a$, Sarah Post$^b$ and Yao-Zhong Zhang$^a$
\newline
\newline
$a.$ School of Mathematics and Physics, The University of Queensland, Brisbane, QLD 4072, Australia
\newline
\newline
$b.$ Department of Mathematics, University of Hawaii at Manoa, Honolulu, HI 96822, USA
\newline
\newline
E-mail: m.hoque@uq.edu.au; i.marquette@uq.edu.au; spost@hawaii.edu;\\ yzz@maths.uq.edu.au
\newline
\newline

\begin{abstract}
We introduce an extended Kepler-Coulomb quantum model in spherical coordinates. The Schr\"{o}dinger equation of this Hamiltonian is solved in these coordinates and it is shown that the wave functions of the system can be expressed in terms of Laguerre, Legendre and exceptional Jacobi polynomials (of hypergeometric type). We construct ladder and shift operators based on the corresponding wave functions and obtain their recurrence formulas. These recurrence relations are used to construct higher-order, algebraically independent integrals of motion to prove superintegrability of the Hamiltonian. The integrals form a higher rank polynomial algebra. By constructing the structure functions of the associated deformed oscillator algebras we derive the degeneracy of energy spectrum of the superintegrable system. 
\end{abstract}

\section{Introduction}
Many families of exceptional orthogonal polynomials have been successfully used to construct new superintegrable systems, higher order integrals of motion and higher order polynomial algebras \cite{Post1, Ian1, Ian2, Ian3, Ian4}. In this paper, we use the recurrence approach to extend the three parameters Kepler-Coulomb system \cite{Kal1}.  

The exceptional orthogonal polynomials (EOP) were first explored in \cite{Kam1, Kam2}. These polynomials form complete, orthogonal systems extending the classical orthogonal polynomials of Hermite, Laguerre and Jacobi. More recently much research has been done extending the theory of  EOPs in various directions in mathematics and physics, in particular, exactly solvable quantum mechanical problems for describing bound states \cite{Gom1, Dut1, Gra1, Gra2, Lev1, Oda1, Que1, Que2, Ses1} and scattering states \cite{Ho1, Yad1, Yad2, Yad3}, diffusion equations and random processes \cite{Ho2, Ho3, Ho4}, quantum information entropy \cite{Dut2}, exact solutions to Dirac equation \cite{Sch1}, Darboux transformations \cite{Oda1, Que1, Que3, Oda2, Kam3, Sas1, Ho5} and finite-gap potentials \cite{Kam4}. Recent progress has been made constructing systems relating superintegrability and supersymmetric quantum mechanics with exceptional orthogonal polynomials \cite{Post1, Mar6}.

The research for superintegrable systems with second-order integrals in conformally flat spaces started in the mid sixties \cite{Fri1}. Over the last decade the topic of superintegrability has become an attractive area of research as these systems possess many desirable properties and can be found throughout various subjects in mathematical physics. For a detailed list of references on superintegrability, we refer the reader to the review paper \cite{Mil1}. One systematic approach to superintegrability is to derive spectra of 2D superintegrable systems based on quadratic and cubic algebras involving three generators \cite{Das1, Mar8, Mar9}. In particular, the method of realization in the deformed oscillator algebras \cite{Das2} has been effective for obtaining finite dimensional unitary representations \cite{Mar8, Mar10}. In fact, this approach was extended to classes of higher order polynomial algebras with three generators \cite{Isa1} as well as higher rank polynomial algebras of superintegrable systems in higher dimensional spaces \cite{FH3, FH4}. However, it is quite involved to apply the direct approach to obtain the corresponding polynomial algebras, Casimir operators and deformed oscillator algebras.

These difficulties can be overcome using a constructive approach based on eigenfunctions of the models. This approach is a useful tool to construct well-defined integrals of motion in classical and quantum mechanical problems. Many papers were devoted to construct integrals of motion and their corresponding higher order symmetry algebras based on lower-(first and second) ones \cite{Fri1, Jau1, Boy1, Eva1, Mar1} and higher-order ladder operators \cite{Ian3, Ian4, Kre1, Adl1, Jun1, Dem1, Mar2, Rag1, Mar4} in various aspects. In fact, the constructive approach has shown a close connection with special functions and (exceptional) orthogonal polynomials \cite{Post1, Mar6, Kal2, Cal1, Cal2, FH1, FH2}.

In this paper, we introduce a new exactly solvable Hamiltonian system in 3D, which is a singular deformation of the Coulomb potential. Its wave functions are given as products of Laguerre, Legendre and exceptional Jacobi polynomials. We show that the system is superintegrable by constructing integrals of the motion using the recurrence relation approach. The symmetry algebra enables us to give an algebraic derivation for the energy spectrum. 

The paper is organized as follows: in section 2, we present a new Hamiltonian system in 3D and show that its Schr\"{o}dinger wave functions can be expressed in terms of Laguerre, Legendre and exceptional polynomials and obtain its physical spectra. In section 3, we construct a set of ladder and shift operators based on the wave functions and show that their suitable combination give the integrals of motion, thus proving the superintegrability of the model. We present the higher rank polynomial algebra generated by these integrals and the realization of this symmetry algebra in terms of the deformed oscillator algebra. By constructing finite-dimensional unitary representation of the symmetry algebra, we obtain the energy spectrum of superintegrable system.

\section{Extended Kepler-Coulomb system}
Consider the generalization of the three parameter Kepler-Coulomb Hamiltonian \cite{Kal1} in the spherical coordinates 
\begin{eqnarray}
&&H=\frac{1}{2}{\bf p}^2 -\frac{\alpha}{2r}+\frac{1}{2r^2\sin^2\theta}\left[\frac{\gamma^2-\frac{1}{4}}{4\sin^2\frac{\phi}{2}}+\frac{\delta^2-\frac{1}{4}}{4\cos^2\frac{\phi}{2}} +\frac{2(1-b\cos\phi)}{(b-\cos\phi)^2}\right],\label{Nh1}
\end{eqnarray}
where $p_i=-i\partial_i$, $b=\frac{\delta+\gamma}{\delta-\gamma}$, $\gamma\neq \delta$ and $\alpha, \gamma, \delta$ are three real constants. 
The Schr\"{o}dinger equation $H\Psi(r,\theta,\phi)=E\Psi(r,\theta,\phi)$ of (\ref{Nh1}) can be expressed as 
\begin{eqnarray}
&&\left[\frac{\partial^2}{\partial r^2}+\frac{2}{r}\frac{\partial}{\partial r}+\frac{\alpha}{r}+2E+\frac{1}{r^2}\left\{\frac{\partial^2}{\partial\theta^2}+\cot\theta\frac{\partial}{\partial\theta}  \right\}\nonumber\right.\\&&\left.+ \frac{1}{r^2\sin^2\theta}\left\{ \frac{\partial^2}{\partial\phi^2} -\frac{\gamma^2-\frac{1}{4}}{4\sin^2\frac{\phi}{2}} -\frac{\delta^2-\frac{1}{4}}{4\cos^2\frac{\phi}{2}}-\frac{2(1-b\cos \phi)}{(b-\cos\phi)^2}\right\} \right]\Psi(r,\theta,\psi)=0.
\end{eqnarray}
The separation of variable of the Hamiltonian (\ref{Nh1}) for the wave equation $H\Psi=E\Psi$ by the ansatz
\begin{eqnarray}
\Psi(r,\theta,\phi)=R(r)\Theta(\theta)Z(\phi)
\end{eqnarray}
provides the following radial and angular ordinary differential equations
\begin{eqnarray}
&&\left[\frac{d^2}{dr^2}+\frac{2}{r}\frac{d}{dr}+\frac{\alpha}{r}+2E-\frac{k_2}{r^2}\right]R(r)=0, \label{Ra1}
\\&&
\left[\frac{d^2}{ d\theta^2}+\cot\theta\frac{d}{d\theta}-\frac{k_1}{\sin^2\theta} +k_2\right]\Theta(\theta)=0,\label{JP1}
\\&&
\left[\frac{d^2}{ d\phi^2} -\frac{\gamma^2-\frac{1}{4}}{4\sin^2\frac{\phi}{2}} -\frac{\delta^2-\frac{1}{4}}{4\cos^2\frac{\phi}{2}}-\frac{2(1-b\cos \phi)}{(b-\cos \phi)^2}+k_1 \right] Z(\phi)=0,\label{EP2}
\end{eqnarray}
where $k_1$, $k_2$ are the associated separation constants.

We now turn to (\ref{EP2}), which can be converted, by setting $z=\cos\phi$, $Z(z)=(z+1)^{\frac{1}{4}(\delta+2)}(z-1)^{\frac{1}{4}(\gamma+2)}(z-b)^{-1}f(z)$, to 
\begin{eqnarray}
&&(z^2-1)\frac{d^2f(z)}{dz^2}+\left\{\gamma-\delta+(\gamma+\delta+2)z-\frac{2(z^2-1)}{z-b}\right\}\frac{df(z)}{dz}\nonumber\\&& +\left\{\frac{1}{4}(\gamma+\delta+1)^2-k_1+\frac{\gamma-\delta+(\gamma+\delta-1)z}{(b-z)}\right\}f(z)=0.\label{Jb3}
\end{eqnarray}
Comparing (\ref{Jb3}) with exceptional Jacobi differential equation \cite{Kam1},
\begin{eqnarray}
T^{(\eta,\xi)}(Y)=(n-1)(n+\eta+\xi)Y,\quad n\in \mathbb{N},
\end{eqnarray}
where
\begin{eqnarray}
&&T^{(\eta,\xi)}(Y)=(X^2-1)Y''+2A\left(\frac{1-B X}{B-X}\right)\{(X-C)Y'-Y\},
\nonumber\\&&
A=\frac{1}{2}(\xi-\eta), \quad B=\frac{\xi+\eta}{\xi-\eta}, \quad C=B+\frac{1}{A},
\end{eqnarray}
we obtain $\gamma=\xi$, $\delta=\eta$ and the separation constant
\begin{eqnarray}
&& k_1=\left(n+\frac{\gamma+\delta-1}{2}\right)^2.\label{Sp2}
\end{eqnarray}
Hence the solutions of (\ref{Jb3}) are given in terms of the exceptional Jacobi polynomials $\hat{P}^{(\delta,\gamma)}_n$ \cite{Post1, Kam1} as 
\begin{eqnarray}
Z(\phi)\equiv F_n(\gamma,\delta)\frac{(\cos \phi+1)^{\frac{1}{4}(2\delta+1)}(\cos \phi-1)^{\frac{1}{4}(2\gamma+1)}}{(\cos\phi-b)}\hat{P}^{(\delta,\gamma)}_n(\cos\phi).
\end{eqnarray}
These EOPs  are related to the standard Jacobi polynomials $P^{(\delta,\gamma)}_n$ \cite{And1} via 
\begin{eqnarray}
\hat{P}^{(\delta,\gamma)}_n=-\frac{1}{2}(\cos\phi-b)P^{(\delta,\gamma)}_{n-1}+\frac{bP^{(\delta,\gamma)}_{n-1}-P^{(\delta,\gamma)}_{n-2}}{\delta+\gamma+2n-2}.
\end{eqnarray}
Using (\ref{Sp2}) in the angular part (\ref{JP1}), we have
\begin{eqnarray}
\left[\frac{d^2}{ d\theta^2}+\cot\theta\frac{d}{d\theta} -\frac{(n+\frac{\gamma+\delta-1}{2})^2}{\sin^2\theta} +k_2\right]\Theta(\theta)=0.\label{JP2}
\end{eqnarray}
Then (\ref{JP2}) can be converted, by setting $z=\cos\theta$, to 
\begin{eqnarray}
&&\left[(1-z^2)\frac{d^2}{dz^2}-2z\frac{d}{dz}+k_2-\frac{(n+\frac{\gamma+\delta-1}{2})^2}{1-z^2}\right] f(z)=0.\label{Jb5}
\end{eqnarray}
Comparing (\ref{Jb5}) with the Legendre differential equation
\begin{eqnarray}
(1-x^2)y''-2xy'+\left[m(m+1)-\frac{\mu^2}{1-x^2}\right] y=0,
\end{eqnarray}
we obtain the constants
\begin{eqnarray}
k_2=m(m+1),\quad \mu=n+\frac{\gamma+\delta-1}{2}.\label{Sp4}
\end{eqnarray}
Hence the solutions of (\ref{JP1}) are given in terms of the Legendre polynomials $P^\mu_m$ \cite{And1} as 
\begin{eqnarray}
\Theta(\theta)\equiv F_m(\mu)P^{\mu}_m(\cos\theta),
\end{eqnarray}
where $F_m(\mu)$ is a normalization constant and $m, \mu\in \mathbb{Z}$.

Using (\ref{Sp4}), the radial part (\ref{Ra1}) becomes
\begin{eqnarray}
\left[\frac{d^2}{dr^2}+\frac{2}{r}\frac{d}{dr}+\frac{\alpha}{r}+2E-\frac{m(m+1)}{r^2}\right]R(r)=0. \label{Rd1}
\end{eqnarray}
(\ref{Rd1}) can be converted, by setting $z=\varepsilon r$, $R(z)=z^m e^{-\frac{1}{2}z}f_1(z)$ and $\varepsilon^2=-8E$, to 
\begin{eqnarray}
\left[z\frac{d^2}{dz^2}+(2m+2-z)\frac{d}{dz}+\frac{\alpha}{\varepsilon}-m-1\right]f_1(z)=0.\label{Lg1}
\end{eqnarray}
Set 
\begin{eqnarray}
N=\frac{\alpha}{\varepsilon}-m-1.
\end{eqnarray}
Then (\ref{Lg1}) can be identified with the Laguerre differential equation. Hence the solutions of (\ref{Ra1}) are given in terms of the $N$-th order Laguerre polynomial functions $L^\beta_N$ \cite{And1} as
\begin{eqnarray}
R(r)\equiv e^{-\frac{\varepsilon r}{2}}(\varepsilon r)^{m}L^{2m+1}_{N}(\varepsilon r).
\end{eqnarray}
Hence the energy spectrum of the model (\ref{Nh1}), $E=\frac{-\varepsilon^2}{8}$,  is given by
\begin{eqnarray}
E=-\frac{\alpha^2}{8\left(N+m+1\right)^2}, \quad N=1, 2, 3,\dots \label{energy}
\end{eqnarray}
Here $N$ represents the principal quantum number.

\section{Algebraic calculation to the extended Kepler-Coulomb system}
We can rewrite the Hamiltonian of the three parameter Kepler-Coulomb system in the standard way as a sequence of operators corresponding to separation in spherical coordinates (\ref{Nh1}),
\begin{eqnarray}
&&H=\frac{1}{2}\left[\frac{\partial^2}{\partial r^2}+\frac{2}{r}\frac{\partial}{\partial r}+\frac{\alpha}{r}+\frac{L_\theta}{r^2} \right],\label{Nh2}
\end{eqnarray}
where
\begin{eqnarray}
&&L_\theta=\frac{\partial^2}{\partial\theta^2}+\cot\theta\frac{\partial}{\partial\theta}+ \frac{L_\phi}{\sin^2\theta}, \\&&
L_\phi= \frac{\partial^2}{\partial\phi^2} -\frac{\gamma^2-\frac{1}{4}}{4\sin^2\frac{\phi}{2}} -\frac{\delta^2-\frac{1}{4}}{4\cos^2\frac{\phi}{2}}-\frac{2(1-b\cos \phi)}{(b-\cos\phi)^2}.
\end{eqnarray}
Making a slight change in the definition of these operators, 
\begin{eqnarray} H_\theta =1-4L_\theta, \qquad H_\phi=-L_\phi,\end{eqnarray}
leads to the following system of eigenvalue equations, from the previous section, 
\begin{eqnarray} H\Psi=E\Psi,\qquad 
H_{\theta} \Psi=\rho^2 \Psi,\qquad
 H_{\phi}\Psi = \mu^2 \Psi. \end{eqnarray}
 Moreover, these three operators mutually commute, i.e.  $[H_\theta, H]=[H_\phi, H]=[H_\theta, H_\phi]=0$.
The wave functions found in the previous section are then
\begin{eqnarray}
\Psi(r,\theta,\phi)=\psi_N^{\rho}\Theta_{\frac{\rho-1}{2}}^\mu\hat{\mathcal{Z}}_n ,
\end{eqnarray}
\begin{eqnarray}
\text{where}&&\psi_N^{\rho}=e^{-\frac{\varepsilon r}{2}}(\varepsilon r)^{\frac{\rho-1}{2}}L^{\rho}_{N}(\varepsilon r),\quad
\Theta_{\frac{\rho-1}{2}}^{\mu}=P^{\mu}_{\frac{\rho-1}{2}}(\cos\theta),
\nonumber\\&&
\hat{\mathcal{Z}}_n=(\cos\phi-b)^{-1}\sin^{\delta+\frac{1}{2}}\frac{\phi}{2}\cos^{\gamma+\frac{1}{2}}\frac{\phi}{2}\hat{P}_n^{(\delta,\gamma)}(\cos\phi).
 \end{eqnarray}
Here $\varepsilon=2\alpha/(2N+\rho+1)$ and 
\[ \rho= 2m+1, \qquad m=0,1,2 \ldots, \]
\[\mu=n+\frac{\gamma+\delta-1}{2}, \qquad n=1,2,3 \ldots\] As in the previous section, the relation among $E$, $N$ and $\rho$ is the quantization condition (\ref{energy})
\begin{eqnarray}
E=-\frac{\alpha^2}{2\left(2N+\rho + 1\right)^2}.\label{Eg1}
\end{eqnarray}
In the following we will construct additional integrals of motion to prove the superintegrability of the Hamiltonian (\ref{Nh2}). 

\subsection{Ladder and shift operators for the associated Laguerre and Legendre polynomials}
We now search for recurrence operators which preserve the energy $E$. Equation (\ref{Eg1}) shows that $E$ is preserved under either
\begin{eqnarray*}
N\rightarrow N+1,\quad \rho\rightarrow \rho-2 \quad \text{or}\quad N\rightarrow N-1,\quad \rho\rightarrow \rho+2,
\end{eqnarray*}
as well as arbitrary shifts in $n$ (equivalently $\mu$).
We now construct the ladder operators from the associated Laguerre functions, as in \cite{Post1, Kal2},
\begin{eqnarray}
L_N=(\rho+1)\frac{\partial}{\partial r}+\alpha-\frac{1}{2r}(\rho^2-1),
\quad
R_N=(-\rho+1)\frac{\partial}{\partial r}+\alpha-\frac{1}{2r}(\rho^2-1),
\end{eqnarray}
whose action on the corresponding wave functions are given by
\begin{eqnarray}
L_N\psi_N^{\rho}=-\frac{2\alpha}{2N+\rho+1}\psi_{N-1}^{\rho+2}, \quad \quad  
R_N\psi_N^{\rho}=-\frac{2\alpha(N+1)(N+\rho)}{2N+\rho+1}\psi_{N+1}^{\rho-2}.
 \end{eqnarray}
We can also construct lowering and rising differential operators of the $\theta$ related part of separated solution for the associated Legendre functions 
\begin{eqnarray}
L_{\rho}=(1-z^2)\frac{\partial}{\partial z}+\frac{\rho-1}{2}z,
\quad
R_{\rho}=(1-z^2)\frac{\partial}{\partial z}-\frac{\rho+1}{2}z,
\end{eqnarray}
where $z=\cos\theta$, and obtain their action on the corresponding wave functions 
\begin{eqnarray}
L_{\rho}\Theta_{\frac{\rho-1}{2}}^{\mu}=(\mu + \frac{\rho-1}{2})\Theta_{\frac{\rho-3}{2}}^{\mu}, 
\quad
R_{\rho}\Theta_{\frac{\rho-1}{2}}^{\mu}=(\mu - \frac{\rho+1}{2})\Theta_{\frac{\rho+1}{2}}^{\mu}.
 \end{eqnarray}
Both of these pairs of ladder operators are obtained by taking the standard ladder operators of the special functions \cite{And1} and conjugating by the ground state. 
 
\subsection{Ladder and shift operators for the exceptional Jacobi polynomials and associated Legendre polynomials}

Ladder operators for the exceptional Jacobi polynomials can be constructed from ladder operators for the Jacobi polynomials \cite{And1}
\begin{eqnarray}
&&\mathcal{L}_n=\frac{1}{2}(2n+\gamma+\delta)(1-y^2)\frac{\partial}{\partial y}-\frac{1}{2}n\{\gamma-\delta-(2n+\gamma+\delta)y\},
\nonumber\\&&
\mathcal{R}_n=-\frac{1}{2}(2n+\gamma+\delta+2)(1-y^2)\frac{\partial}{\partial y}+\frac{1}{2}(n+\gamma+\delta+1)\nonumber\\&&\qquad\quad\times\{\gamma-\delta+(2n+\gamma+\delta+2)y\}.
\end{eqnarray}
Their action is as
\begin{eqnarray}
&&\mathcal{L}_{n} P_n^{(\delta,\gamma)}(y)=(n+\gamma)(n+\delta) P_{n-1}^{(\delta,\gamma)}(y),
\nonumber\\&&
\mathcal{R}_{n}P_n^{(\delta,\gamma)}(y)=(n+1)(n+\gamma+\delta+1) P_{n+1}^{(\delta,\gamma)}(y),\quad y=\cos \phi.
\end{eqnarray}
To extend these operators to the EOP case, we make use of forward and backward operators  \cite{Post1, Ho5} for the exceptional Jacobi polynomials
\begin{eqnarray}
&&\mathcal{F}=(y-1)(y+\frac{\gamma+\delta}{\delta-\gamma})\frac{\partial}{\partial y}+\delta(y+\frac{2+\gamma+\delta}{\delta-\gamma}),
\nonumber\\&&
\mathcal{B}=\frac{\gamma-\delta}{\gamma+\delta-(\gamma-\delta)y}\{(1+y)\frac{\partial}{\partial y}+\delta\}, \quad y=\cos \phi,
\end{eqnarray}
whose actions are 
\begin{eqnarray}
\mathcal{F} P_n^{(\delta+1,\gamma-1)}(y)=-2(n+\delta-1)\hat{P}_{n+1}^{(\delta,\gamma)}(y),
\\
\mathcal{B}\hat{P}_{n}^{(\delta,\gamma)}(y)=\frac{1}{2}(n+\gamma)  P_{n-1}^{(\delta +1,\gamma-1)}(y).
\end{eqnarray}
We can then define, as in  \cite{Post1}, the corresponding ladder operators for the exceptional Jacobi polynomials via
\begin{eqnarray}
L_n=\mathcal{F}\circ \mathcal{L}^{-}_n\circ \mathcal{B}, \quad R_n=\mathcal{F}\circ \mathcal{R}^{+}_n\circ \mathcal{B}.
\end{eqnarray}
The final step is to conjugate these  ladder operators by the ground state $y_0=(y+1)^{\frac{1}{4}(\delta+2)}(y-1)^{\frac{1}{4}(\gamma+2)}(y-b)^{-1}$ for the angular component of the eigenfunction 
\begin{eqnarray}
L_n=y_0 L_n y_0^{-}, \quad R_n=y_0 R_n y_0^{-},\label{ep1}
\end{eqnarray}
so that their actions on the $\phi$-components of the wave function are as follows,
\begin{eqnarray}
&&L_n \hat{\mathcal{Z}}_{n}=-(n+\delta)(n+\gamma)(n+\delta-2)(n+\gamma-2)\hat{\mathcal{Z}}_{n-1},
\nonumber\\&&
R_n\hat{\mathcal{Z}}_{n}=-n(n+\delta)(n+\gamma)(n+\delta+\gamma-1)\hat{\mathcal{Z}}_{n+1}.
\end{eqnarray}
While these operators shift the parameter $n$ (equivalently $\mu$) in the $\phi$-factor $\hat{\mathcal{Z}}_{n}$, we must account for this shift in the $\Theta^{\mu}_{\frac{\rho-1}{2}}(z)$ component as well. To do so, 
we now construct a pair of operators from associated Legendre polynomials that can lower and rise $\mu$ while fixing $\rho$,
\begin{eqnarray}
L_{\mu}=\sqrt{1-z^2}\frac{\partial}{\partial z}-\frac{\mu z}{\sqrt{1-z^2}},
\quad
R_{\mu}=\sqrt{1-z^2}\frac{\partial}{\partial z}+\frac{\mu z}{\sqrt{1-z^2}},
\end{eqnarray}
where $z=\cos\theta$, and their actions on the corresponding wave functions are given by
\begin{eqnarray}
L_{\mu}\Theta^{\mu}_{\frac{\rho-1}{2}}(z)=(\frac{\rho-1}{2}+\mu)(\frac{\rho+1}{2}-\mu)\Theta^{\mu-1}_{\frac{\rho-1}{2}}(z),
\quad
R_{\mu}\Theta^{\mu}_{\frac{\rho-1}{2}}(z)=-\Theta^{\mu+1}_{\frac{\rho-1}{2}}(z).
\end{eqnarray}

\subsection{Integrals of motion and algebraic structure}
Let us now consider the following suitable combinations of the operators
\begin{eqnarray}
D_1^{-}=L_N R_\rho, \quad D_1^{+}=R_N L_\rho,
\quad
D_2^{-}=R_\mu L_n, \quad D_2^{+}=L_\mu R_n.
\end{eqnarray}
The action of the operators $D^{\pm}_{i}$, $i=1, 2$ fixes our complete basis of eigenfunctions, thus providing  higher order integrals of the motion. Their explicitly action on the eigenfunctions are  given by
\begin{eqnarray}
&D_1^{-}\Psi(r, \theta, \phi)&=\frac{(\rho-2\mu+1)\alpha}{(2N+\rho+1)}\psi_{N-1}^{\rho+2}\Theta_{\frac{\rho+1}{2}}^\mu\hat{\mathcal{Z}}_n,
\nonumber\\&
 D_1^{+}\Psi(r, \theta, \phi)&=-\frac{\alpha(N+1)(N+\rho)(\rho+2\mu -1)}{2N+\rho+1}\psi_{N+1}^{\rho-2}\Theta_{\frac{\rho-3}{2}}^\mu\hat{\mathcal{Z}}_n,
\\
& D_2^{-}\Psi(r, \theta, \phi)&=(n+\delta)(n+\gamma)(n+\delta-2)(n+\gamma-2) \psi_{N}^{\rho}\Theta_{\frac{\rho-1}{2}}^{\mu+1}\hat{\mathcal{Z}}_{n-1},\nonumber\\
&D_2^{+}\Psi(r, \theta, \phi)&=\frac{-1}{4}n(n+\delta)(n+\gamma)(n+\delta+\gamma-1)(\rho+2\mu-1)\nonumber\\&&\quad\times(\rho-2\mu+1) \psi_{N}^{\rho}\Theta_{\frac{\rho-1}{2}}^{\mu-1}\hat{\mathcal{Z}}_{n+1}.
\end{eqnarray} 
The following commutation relations of the operators can be easily verified via the action on the eigenfunctions  (\ref{Nh2}),
\begin{eqnarray}
&&[D_1^{-}, H]=0 = [D_1^{+}, H],\quad
[D_1^{-}, H_\phi]=0=[D_1^{+}, H_\phi],
\nonumber\\&&
[D_2^{-}, H]=0=[D_2^{+}, H],
\quad
[D_2^{-}, H_\theta]=0=[D_2^{+}, H_\theta].
\end{eqnarray}
For the convenience we present a diagram representation of the above commutation relations
\begin{eqnarray}
\begin{xy}
(0,0)*+{D^{-}_{1}}="m";(20,0)*+{H_{\phi}}="f"; (40,0)*+{D^{+}_{1}}="n"; (20,20)*+{H}="r";   (60,0)*+{D^{-}_{2}}="j";(80,0)*+{H_{\theta}}="k"; (100,0)*+{D^{+}_{2}}="l"; (80,20)*+{H}="g"; 
"f";"r"**\dir{--};
"j";"k"**\dir{--}; 
"l";"k"**\dir{--};
"l";"g"**\dir{--};
"g";"j"**\dir{--};
"g";"k"**\dir{--};
"f";"n"**\dir{--}; 
"m";"f"**\dir{--};
"n";"r"**\dir{--};
"r";"m"**\dir{--};
\end{xy}
\end{eqnarray}
Moreover, we obtain
\begin{eqnarray}
&&[H_\theta, D_1^{-}]=\frac14(\rho+1)D_1^{-},\quad\quad \quad \quad [H_\phi, D_2^{-}]=(2\mu+1) D_2^{-},
\nonumber\\&&
[H_\theta, D_1^{+}]=\frac{-1}{4}(\rho-1) D_1^{+}, \quad\quad\quad\quad [H_\phi, D_2^{+}]= -(2\mu-1)D_2^{+}.
\end{eqnarray}
Let us now define the higher order operators $D^{\mp}_{i}D^{\pm}_{i}$, $i=1, 2$. We can also obtain the action of the operators $D^{\mp}_{i}D^{\pm}_{i}$, $i=1, 2$ on the wave functions. It follows from the construction that  they are algebraically independent sets of differential operators and hence the system is superintegrable. The system also evidences  a common feature of superintegrable systems in that is admits higher-order algebraic structure. A direct computation of the action of the operators $D_1^{\pm}$ on the basis leads to 
\begin{eqnarray}
[H_\theta, D_1^{-}]=\frac14(\sqrt{H_\theta}+1)D_1^{-}, \quad [H_\theta, D_1^{+}]=-\frac{-1}{4}(\sqrt{H_\theta}-1) D_1^{+},\label{d1}
\end{eqnarray}
\begin{eqnarray}
&D_1^{-}D_1^{+}& =\frac{1}{4}[\sqrt{-\alpha^2}-\sqrt{2H}\sqrt{H_\theta}+\sqrt{2H}][\sqrt{-\alpha^2}+\sqrt{2H}\sqrt{H_\theta}-\sqrt{2H}]\nonumber\\&&\times [\sqrt{H_\theta}+2\sqrt{-L_\phi}-1][\sqrt{H_\theta}-2\sqrt{H_\phi}-1],
\nonumber\\&
D_1^{+}D_1^{-} &=\frac{1}{4}[\sqrt{-\alpha^2}-\sqrt{2H}\sqrt{H_\theta}-\sqrt{2H}][\sqrt{-\alpha^2}+\sqrt{2H}\sqrt{H_\theta}+\sqrt{2H}]\nonumber\\&&\times [\sqrt{H_\theta}+2\sqrt{H_\phi}+1][\sqrt{H_\theta}-2\sqrt{H_\phi}+1].\label{d2}
\end{eqnarray}
Similarly for the $D_2^{\pm}$ operators
\begin{eqnarray}
[H_\phi, D_2^{-}]=(2\sqrt{H_\phi}+1) D_2^{-}, \quad [ H_\phi, D_2^{+}]= -(2\sqrt{H_\phi}-1)D_2^{+},\label{d3}
\end{eqnarray}
\begin{eqnarray}
&D_2^{-}D_2^{+}&  =-\frac{1}{1024}[1 + \sqrt{H_\theta}+2 \sqrt{H_\phi}][-1 +\sqrt{H_\theta} - 2\sqrt{H_\phi}]\nonumber\\&&\times [1 + 2\sqrt{H_\phi} - \gamma - \delta][-1 + 2\sqrt{H_\phi} + \gamma - \delta] [1 + 2\sqrt{H_\phi} + \gamma - \delta]\nonumber\\&&\times[3 + 2\sqrt{H_\phi} + \gamma - \delta][-1 + 2\sqrt{H_\phi} - \gamma + \delta][1 + 2\sqrt{H_\phi} - \gamma + \delta]\nonumber\\&&\times [3 + 2\sqrt{H_\phi} - \gamma + \delta][-1 + 2\sqrt{H_\phi} + \gamma + \delta],\nonumber\\
&D_2^{+}D_2^{-} &=-\frac{1}{1024}[-1 + \sqrt{H_\theta} {+}2 \sqrt{H_\phi}][1 + \sqrt{H_\theta} {-} 2 \sqrt{H_\phi}]\nonumber\\&&\times [-1 + 2 \sqrt{H_\phi} - \gamma - \delta][-3 + 2 \sqrt{H_\phi} + \gamma - \delta][-1 +2\sqrt{H_\phi} + \gamma - \delta]  \nonumber\\&&\times [1 + 2\sqrt{H_\phi} + \gamma - \delta][-3 + 2 \sqrt{H_\phi} - \gamma + \delta][-1 + 2\sqrt{H_\phi} - \gamma + \delta] \nonumber\\&&\times [1 + 2\sqrt{H_\phi} - \gamma + \delta][-3 + 2\sqrt{H_\phi} + \gamma + \delta].\label{d4}
\end{eqnarray}
So the above higher order algebraic structure is the full symmetry algebra for the superintegrable system (\ref{Nh2}).
\subsection{Higher rank polynomial algebra}
In this subsection we will redefine the operators in sense of \cite{Kal1} and show that they form a well-defined higher rank polynomial algebra. 
We now define the following operators as 
\begin{eqnarray}
&&J_1=\frac{D^{-}_1-D^{+}_1}{\rho}, \quad  J_2=D^{-}_1+D^{+}_1,
\nonumber\\&&
K_1=\frac{D^{+}_2-D^{-}_2}{2\mu}, \quad  K_2=D^{-}_2+D^{+}_2.
\end{eqnarray}
It is easily verified that 
\begin{eqnarray}
&&[J_1, H]=0 = [J_2, H],\quad\quad [J_1, H_\phi]=0=[J_2, H_\phi],
\nonumber\\&&
[K_1, H]=0 = [K_2, H],\quad [K_1, H_\theta]=0=[K_2, H_\theta].
\end{eqnarray}
The commutation relations also can be represented by the following diagrams
\begin{eqnarray}
\begin{xy}
(0,0)*+{J_{1}}="m";(20,0)*+{H_{\phi}}="f"; (40,0)*+{J_{2}}="n"; (20,20)*+{H}="r";   (60,0)*+{K_{1}}="j";(80,0)*+{H_{\theta}}="k"; (100,0)*+{K_{2}}="l"; (80,20)*+{H}="g"; 
"f";"r"**\dir{--};
"j";"k"**\dir{--}; 
"l";"k"**\dir{--};
"l";"g"**\dir{--};
"g";"j"**\dir{--};
"g";"k"**\dir{--};
"f";"n"**\dir{--}; 
"m";"f"**\dir{--};
"n";"r"**\dir{--};
"r";"m"**\dir{--};
\end{xy}
\end{eqnarray}
We obtain the following, still quantum-number dependent,  commutation relations
\begin{eqnarray}
&&[H_\theta, J_1]=\frac{1}{4}\left(J_1+J_2\right), \quad [ H_\theta, J_2]=\frac{1}{4}\left(\rho^2 J_1+J_2\right),
\nonumber\\&&
[H_\phi, K_1]=K_1+K_2, \quad [ H_\phi, K_2]=(2n+\delta+\gamma-1)^2 K_1+K_2.
\end{eqnarray}
These can be expressed back in terms of the algebra generators as  
\begin{eqnarray}
&&[H_\theta, J_1]=\frac{1}{4}\left(J_1+J_2\right), \quad [ H_\theta, J_2]=\frac{1}{4}\left(H_\theta J_1+J_2\right),
\nonumber\\&&
[H_\phi, K_1]=K_1+K_2, \quad [ H_\phi, K_2]=4 H_\phi K_1+K_2.
\end{eqnarray}
The last set of algebra relations to recover are the commutators $[J_1, J_2]$ and $[K_1, K_2].$ A first step is to see the following relations from the action on the eigenfunctions
\begin{eqnarray}
&&[J_1, J_2]=\frac{2}{\sqrt{H_\theta}}[D_1^{-}, D_1^{+}],
\quad
[K_1, K_2]=\frac{1}{\sqrt{H_\phi}}[D_2^{+}, D_2^{-}].
\end{eqnarray}
Moreover, $[J_1,K_1]=0=[J_2,K_2]$ as $[D_1^{\pm},D_2^{\pm}]=0.$
We can rewrite the expressions (\ref{d2}) as 
\begin{eqnarray}
&D_1^{-}D_1^{+}& =P_1(H, H_\theta, H_\phi)+P_2(H, H_\theta, H_\phi)\sqrt{H_\theta},
\nonumber\\&
D_1^{+}D_1^{-} &=P_1(H, H_\theta, H_\phi)-P_2(H, H_\theta, H_\phi)\sqrt{H_\theta},
\end{eqnarray}
where
\begin{eqnarray}
&P_1(H, H_\theta, H_\phi)&=-\frac{1}{4}[16 H  - 64 H H_\theta  + 32 H H_\theta^2 + 16 H H_\phi  - 32 H H_\theta H_\phi \nonumber\\&& \quad +  2 \alpha^2  - 4 H_\theta \alpha^2 + 4 H_\phi \alpha^2],
\nonumber\\&
P_2(H,H_\theta, H_\phi)& =\frac{1}{4}[ 16 H  - 32 H  H_\theta + 16 H  H_\phi  + 2 \alpha^2 ].
\end{eqnarray}
Hence we have 
\begin{eqnarray}
&&[D^{-}_1, D^{+}_1]= 2P_2(H,H_\theta, H_\phi)\sqrt{H_\theta},
\\&&
 \{D^{-}_1, D^{+}_1\}= 2P_1(H,H_\theta, H_\phi). 
\end{eqnarray}
Also, the expressions (\ref{d4}) can be written as
\begin{eqnarray}
&D_2^{-}D_2^{+}& =P_3(H_\theta, H_\phi)+P_4(H_\theta, H_\phi)\sqrt{H_\phi},
\nonumber\\&
D_2^{+}D_2^{-} &=P_3(H_\theta, H_\phi)-P_4(H_\theta, H_\phi)\sqrt{H_\phi},
\end{eqnarray}
where 
\begin{eqnarray}
&P_3(H_\theta, H_\phi)&=-\frac{1}{1024}[((\gamma+\delta-1)^2 - 4 H_\phi][\{ (\gamma-\delta)^2-9\}(H_\theta-1)\nonumber\\&&\qquad - 4 \{(\gamma-\delta)^2 + H_\theta-22\}H_\phi + 16 H_\phi^2][(\gamma-\delta)^4 +(4 H_\phi-1)^2 \nonumber\\&&\qquad - 2 (\gamma-\delta)^2 (4 H_\phi+1)],
\nonumber\\&
P_4(H_\theta, H_\phi)& =\frac{1}{256} [(\gamma+\delta-1)^2 - 4 H_\phi] [(\gamma-\delta)^4 + (1 - 4 H_\phi)^2 - 2 (\gamma-\delta)^2 (1 + 4 H_\phi)]\nonumber\\&&\qquad \times [(\gamma-\delta)^2 + 3 H_\theta - 4 (3 + 4 H_\phi)].
\end{eqnarray}
Hence we also have 
\begin{eqnarray}
&&[D^{-}_2, D^{+}_2]= 2P_4(H_\theta, H_\phi)\sqrt{H_\phi},
\\&&
 \{D^{-}_2, D^{+}_2\}= 2P_3(H_\theta, H_\phi). \label{algdep}
\end{eqnarray}
Thus, we realize the final set of algebra relations as 
\begin{eqnarray}
&&[J_1, J_2]=4P_2(H,H_\theta, H_\phi),
\quad
[K_1, K_2]=2P_4(H_\theta, H_\phi).
\end{eqnarray}

Thus, we have shown that the operators $H$, $H_{\theta}$, $H_\phi$, $K_1$, $K_2,$ $J_1$ and $J_2$ close to form a higher-rank polynomial algebra. Finally, we mention that it is possible to show that these operators are well-defined and can be expressed without recourse to the action on the wave-functions. This is accomplished via the usual observation that the operators constructed are polynomial in $\rho^2$ and $\mu^2$ and so these can be replaced with the appropriate operators and the algebra relations will still hold. 

\subsection{Deformed oscillators, structure functions and spectrum}
 In order to derive the spectrum using the algebraic structure, we realize the substructure (\ref{d1}) and (\ref{d2}) as well as the substructure (\ref{d3}) and (\ref{d4}), respectively, in terms of deformed oscillator algebra \cite{Das1,Das2} $\{\aleph, b^{\dagger}, b\}$ of the form
\begin{eqnarray}
[\aleph,b^{\dagger}]=b^{\dagger},\quad [\aleph,b]=-b,\quad bb^{\dagger}=\Phi (\aleph+1),\quad b^{\dagger} b=\Phi(\aleph),\label{kpfh}
\end{eqnarray}
where $\aleph $ is the number operator and $\Phi(x)$ is well behaved real function satisfying 
\begin{eqnarray}
\Phi(0)=0, \quad \Phi(x)>0, \quad \forall x>0.\label{kpbc}
\end{eqnarray}
We recall (\ref{d1}) and (\ref{d3}) in the following forms
\begin{eqnarray}
[\sqrt{H_\theta}, D_1^{\mp}]=\pm 2D_1^{\mp}, \quad
[\sqrt{H_\phi}, D_2^{\pm}]=\pm D_2^{\pm}.
\end{eqnarray}
Setting
\begin{eqnarray}
&&\sqrt{H_\theta}=2(\aleph_1+u_1), \quad b_1=D_1^{+}, \quad b_1^{\dagger}=D_1^{-},
\\&&\sqrt{H_\phi}=(\aleph_2+u_2), \quad b_2=D_2^{-}, \quad b_2^{\dagger}=D_2^{+},
\end{eqnarray}
where  $u_1>0$ and $u_2>0$ are some representation dependent constants, we obtain from (\ref{d2}) and (\ref{d4}),
\begin{eqnarray}
&&[\aleph_i,b^{\dagger}_i]=b^{\dagger}_i,\quad\quad [\aleph_i,b_i]=-b_i, \quad i=1,2, \nonumber\\
&& b_1b^{\dagger}_1=\Phi_1 (\aleph_1+1,\aleph_2+1, H, u_1, u_2),\quad b^{\dagger}_1 b_1=\Phi_1 (\aleph_1,\aleph_2,H, u_1, u_2),\nonumber\\
&& b_2b^{\dagger}_2=\Phi_2 (\aleph_1+1,\aleph_2+1, u_1, u_2),\quad b^{\dagger}_2 b_2=\Phi_1 (\aleph_1,\aleph_2, u_1, u_2).
\label{kpfh1}
\end{eqnarray}
The corresponding structure functions are given by
\begin{eqnarray}
\Phi_1(\aleph_1,\aleph_2, H, u_1, u_2)&=& b_1^{\dagger} b_1=D^{-}_1D^{+}_1 \nonumber\\
&=& P_1(\aleph_1,\aleph_2, H, u_1, u_2)+2 P_2(\aleph_1,\aleph_2, H, u_1, u_2)(\aleph_1+u_1),\nonumber\\
\end{eqnarray}
\begin{eqnarray}
\Phi_2(\aleph_1,\aleph_2, u_1, u_2)&=& b_2^{\dagger} b_2=D^{+}_2D^{-}_2 \nonumber\\
&=& P_3(\aleph_1,\aleph_2, u_1, u_2)+2 P_4(\aleph_1,\aleph_2, u_1, u_2)(\aleph_2+u_2).
\end{eqnarray}
At this stage, the explicit expressions of the corresponding structure functions are as follows
\begin{eqnarray}
\Phi_1(\aleph_1,\aleph_2, E, u_1, u_2) &=&\frac{1}{4}\left[\sqrt{-\alpha^2}-2(\aleph_1+u_1)\sqrt{2E}+\sqrt{2E}\right]\nonumber\\
& &\left[\sqrt{-\alpha^2}+2(\aleph_1+u_1)\sqrt{2E}-\sqrt{2E}\right]\nonumber\\ 
& &[2(\aleph_1+u_1)+2(\aleph_2+u_2)-1]\nonumber\\
& &[2(\aleph_1+u_1)-2(\aleph_2+u_2)-1],
\end{eqnarray}
\begin{eqnarray}
\Phi_2(\aleph_1,\aleph_2, u_1, u_2) &=&-\frac{1}{1024}\left[-1 + 2(\aleph_1+u_1) + 2(\aleph_2+u_2)\right]\nonumber\\
& & [1 + 2(\aleph_1+u_1)- 2 (\aleph_2+u_2)]\, [-1 + 2 (\aleph_2+u_2) - \gamma - \delta]\nonumber\\
& &[-3 + 2 (\aleph_2+u_2) + \gamma - \delta] \,[-1 +2(\aleph_2+u_2) + \gamma - \delta] \nonumber\\
& & [1 + 2(\aleph_2+u_2) + \gamma - \delta] \,
 [-3 + 2 (\aleph_2+u_2) - \gamma + \delta]\nonumber\\
& & [-1 + 2(\aleph_2+u_2) - \gamma + \delta] \, 
 [1 + 2(\aleph_2+u_2) - \gamma + \delta]\nonumber\\
& & [-3 - 2(\aleph_2+u_2) + \gamma + \delta].
\end{eqnarray}
Note that only $\Phi_1$ contains the energy parameter $E$. To determine the energy spectrum, we need to construct the finite-dimensional unitary representations of (\ref{kpfh1}). We thus impose the following constraints on the structure functions:
\begin{eqnarray}
&&\Phi_1(p_1+1, p_2+1, E, u_1, u_2)=0,\quad \Phi_1(0,0, E, u_1, u_2)=0,\label{Cont1}\\
&&\Phi_2(p_1+1, p_2+1, u_1, u_2)=0,\quad \Phi_2(0,0, u_1, u_2)=0,\label{Cont2}
\end{eqnarray}
where $p_i, i=1,2$, are positive integers. These constraints give rise to finite-dimensional unitary presentations. We now solve the constraints (\ref{Cont1}) and (\ref{Cont2}) simultaneously. 
First of all, it can be readily verified that the only solution for the constraints ((\ref{Cont2}) is given by
\begin{eqnarray}
u_2=u_1+\frac{1}{2},\quad\quad p_1=p_2=p,\label{Soln-to-Phi2}
\end{eqnarray}
where $p$ is a positive integer. It follows that these constraints (\ref{Cont1}) and (\ref{Cont2}) lead to $(p+1)$-dimensional unitary representations of (\ref{kpfh1}).  Now we find solutions to the constraints (\ref{Cont1}) which satisfy (\ref{Soln-to-Phi2}). This will provide us the energy spectrum $E$ of the system as well as the allowed values of the parameters $u_1$ and $u_2$. After some computations, we obtain all allowed values of the energy $E$ and the parameters $u_1$, $u_2$ as follows.
\begin{eqnarray}
 E=-\frac{\alpha^2}{2(p+1)^2} ,\quad u_1=\frac{1}{2}+\frac{p+1}{2}, \quad u_2=1+\frac{p+1}{2}.
\end{eqnarray}
Here $p=0,1, \cdots,$ is any positive integer. Making the identification $p=2(N+m)+1$, the energy spectrum becomes (\ref{energy}).

\section{Conclusion}
In this paper, we see the construction of a new, exactly-solvable system with wave functions comprised of products of Laguerre, Legendre and exceptional Jacobi polynomials. This system is a perturbation of a superintegrable system, the singular Coulomb system. We show that the system is superintegrable by constructing 7 integrals of motion and show that the algebra closes to form a polynomial algebra.  The construction of the higher order integrals of the motion is a systematic constructive approach from ladder operators which based on (exceptional) orthogonal polynomials. We also discuss the representations of this new algebra via the deformed oscillator method to derive spectra and degeneracies of unitary representations. 

It would be of interest to further investigate this system further, in particular to understand how the singular term behaves in the classical limit. Here we have normalized $\hbar=1$, if it is reintroduced the system will depend non-trivially on this parameter. Also of interest could be to understand the scattering states and how the exception perturbation affects those. 

{\bf Acknowledgements:} The research of FH was supported by International Postgraduate Research Scholarship and Australian Postgraduate Award. He would like to thank the Graduate School, UQ for supporting travel grant of this project, and Dr. Sarah Post for her great hospitality in Hawaii. IM was supported by the Australian Research Council, Discovery Project DP 160101376, YZZ was partially supported by the Australian Research Council, Discovery Project DP 140101492, SP was partially supported by the Simon's Foundation Grant \# 319211.


\begin{thebibliography}{}

\bibitem{Post1} Post S, Tsujimoto and Vinet L  2012 Families of superintegrable Hamiltonians constructed from exceptional polynomials \textit{J. Phys. A: Math. Theor.} {\bf 45}, 405202.

\bibitem{Ian1} Marquette I and Quesne  2013 New families of superintegrable systems from Hermite and Laguerre exceptional orthogonal polynomials \textit{J. Math. Phys.} {\bf 54}, 042102.

\bibitem{Ian2} Marquette I and Quesne  2013 Two-step rational extensions of the harmonic oscillator: exceptional orthogonal polynomials and ladder operators \textit{J. Phys. A: Math. Theor.} {\bf 46}, 155201.

\bibitem{Ian3} Marquette I and Quesne  2013 New ladder operators for a rational extension of the harmonic oscillator and superintegrability of some two-dimensioanl systems \textit{J. Math. Phys.} {\bf 54}, 102102.

\bibitem{Ian4} Marquette I and Quesne  2014 Combined state-adding and state-deleting approaches to type $III$ multi-step rationally extended potentials: Applications to ladder operators and superintegrability \textit{J. Math. Phys.} {\bf 55}, 112103.

\bibitem{Kal1} Kalnins E G, Kress J M and Miller W Jr 2013 Extended Kepler-Coulomb quantum superintegrable systems in three dimensions  \textit{J. Phys. A: Math. Theor.} {\bf 46}, 085206.

\bibitem{Kam1} G\'{o}mez-Ullate D, Kamran N and Milson R 2009 An extended class of orthogonal polynomials defined by a Sturm-Liouville problem \textit{J. Math. Anal. Appl.} {\bf 359}, 359.

\bibitem{Kam2} G\'{o}mez-Ullate D, Kamran N and Milson R 2010 An extension of Bochner's problem: exceptional invariant subspaces  \textit{J. Appro. Theor.} {\bf 162}, 987.

\bibitem{Gom1} G\'{o}mez-Ullate D, Grandati Y and Milson R 2014 Rational extensions of the quantum harmonic oscillator and exceptional hermit polynomials \textit{J. Phys. A: Math. Theor.} {\bf 47}, 015203.

\bibitem{Dut1} Dutta D and Roy P 2010 Conditionally exactly solvable potentials and exceptional orthogonal polynomials \textit{J. Math. Phys.} {\bf 51}, 042101.

\bibitem{Gra1} Grandati Y 2011 Solvable rational extensions of the isotonic oscillator \textit{Annal. of Phys.} {\bf 326}, 2074.

\bibitem{Gra2} Grandati Y 2011 Solvable rational extensions of the Morse and Kepler-Coulomb potentials \textit{J. Math. Phys.} {\bf 52}, 103505.

\bibitem{Lev1} L\'{e}vai G and \"{O}zer O 2010 An exactly solvable Schr\"{o}dinger equation with finite positive position-dependent effective mass \textit{J. Math. Phys.} {\bf 51}, 092103.

\bibitem{Oda1} Odake S and Sasaki R 2009 Infinitely many shape invariant potentials and new orthogonal polynomials \textit{ Phys. Lett. B} {\bf 679}, 414.

\bibitem{Que1} Quesne C 2011 Higher-order SUSY, exactly solvable potentials, and exceptional orthogonal polynomials \textit{ Mod. Phys. Lett. A} {\bf 26}, 1843.

\bibitem{Que2} Quesne C 2009 Solvable rational potentials and exceptional orthogonal polynomials in supersymmetric quantum mechanics, symmetry, integrability and geometry: methods and applications \textit{ SIGMA} {\bf 5}.

\bibitem{Ses1} Sesma J 2010 The generalize quantum isotonic oscillator \textit{J. Phys. A: Math. Theor.} {\bf 43}, 185303.

\bibitem{Ho1} Ho C L, Lee J C and Sasaki R 2014 Scattering amplitudes for multi-indexed extensions of solvable potentials \textit{Annal. Phys.} {\bf 343}, 115.

\bibitem{Yad1} Yadav R K, Khare A and Mandal B P 2015 The scattering amplitude for rationally extended shape invariant Eckart potentials \textit{Phys. Lett. A} {\bf 379}, 67.

\bibitem{Yad2} Yadav R K, Khare A and Mandal B P 2013 The scattering amplitude for a newly found exactly solvable potential \textit{Annal. Phys. } {\bf 331}, 313.

\bibitem{Yad3} Yadav R K, Khare A and Mandal B P 2013 The scattering amplitude for one parameter family of shape invariant potentials related to Jacobi polynomials \textit{Phys. Lett. B } {\bf 723}, 433.

\bibitem{Ho2} Ho C L 2011 Direc (-Pauli), Fokker-Planck equations and exceptional Laguerre polynomials \textit{Annal. Phys.} {\bf 326}, 797.

\bibitem{Ho3} Ho C L and Sasaki R 2014 Extensions of a class of similarity solutions of Fokker-Planck equation with time-dependent coefficients an fixed/moving boundaries \textit{J. Math. Phys.} {\bf 55}, 113301.

\bibitem{Ho4} Chou C I and Ho C L 2013 Generalized Rayleigh and Jacobi processes and exceptional orthogonal polynomials \textit{Int. J. Mod. Phys. B} {\bf 27}, 1350135.

\bibitem{Dut2} Dutta D and Roy P 2011 Information entropy of conditionally exactly solvable potentials \textit{J. Math. Phys.} {\bf 52}, 032104.

\bibitem{Sch1} Schulze-Halberg A and Roy B 2014 Darbouz partners of pseudoscalar Dirac potentials associated with exceptional orthogonal polynomials \textit{Annal.  Phys.} {\bf 349}, 159.

\bibitem{Que3} Quesne C 2008 Exceptional orhtogonal polynomials, exactly solvable potentials and supersymmetry \textit{ J. Phys. A: Math. Theor.} {\bf 41}, 392001.

\bibitem{Oda2} Odake S and Sasaki R 2010 Another set of infinitely many exceptional $X_l$ Laguerre polynomials \textit{ Phys. Lett. B} {\bf 684}, 173.

\bibitem{Kam3} G\'{o}mez-Ullate D, Kamran N and Milson R 2010 Exceptional orthogonal polynomials and the Darboux transformation \textit{J. Phys. A: Math. Theor.} {\bf 43}, 434016.

\bibitem{Sas1} Sasaki R, Tsujimoto S and Zhedanov 2010 Exceptional Laguerre and Jacobi polynomials and the corresponding potentials through Darboux-Crum transformations \textit{J. Phys. A: Math. Theor.} {\bf 43}, 315204.

\bibitem{Ho5} Ho C L, Odake S and Sasaki R 2011 Properties of the exceptional ($X_l$) Laguerre and Jacobi polynomials \textit{SIGMA} {\bf 7}, 107.

\bibitem{Kam4} G\'{o}mez-Ullate D, Kamran N and Milson R 2012 Two step Darboux transformations and exceptional Laguerre polynomials \textit{J. Math. Anal, Appl.} {\bf 387}, 410.

\bibitem{Mar6}  Marquette I and Quesne C  2015 Deformed oscilator algebra approach of some quantum superintegrable Lissajous systems on the sphere and of their rational extensions \textit{J. Math. Phys.} {\bf 56}, 062102.

\bibitem{Fri1} Fris I, Smorodinsky Y A, Uhlir M and Winternitz P 1966 Symmetry groups in classical and quantum mechanics \textit{Yad Fiz} {\bf 4,} 625 (Sov. J. Nucl. Phys. {\bf 4,} 444).

\bibitem{Mil1} Miller W Jr, Post S and Winternitz P 2013 Classical and quantum superintegrability with applications \textit{J. Phys. A: Math. Theor.} {\bf 46,} 423001.

\bibitem{Das1}  Daskaloyannis C 2001 Quadratic Poisson algebras of two-dimensional classical superintegrable systems and quadratic associative algebras of quantum superintegrable systems \textit{J. Math. Phys.} {\bf 42}, 1100.

\bibitem{Mar8} Marquette I 2009 Superintegrability with third order integrals of motion, cubic algebras and supersymmetric quantum mechanics I:Rational function potentials \textit{J. Math. Phys.} {\bf 50}, 012101.

\bibitem{Mar9} Marquette I 2009 Superintegrability with third order integrals of motion, cubic algebras and supersymmetric quantum mechanics II:Painleave transcendent potentials \textit{J. Math. Phys.} {\bf 50}, 095202.

\bibitem{Das2}  Daskaloyannis C 1991 Generalized deformed oscillator and nonlinear algebras \textit{J. Phys. A: Math. Gen.} {\bf 24}, L789.

\bibitem{Mar10} Marquette I 2013 Quartic Poisson algebras and quartic associative algebras and realizations as deformed oscillator algebras \textit{J. Math. Phys.} {\bf 54}, 071702.

\bibitem{Isa1} Isaac P S and Marquette I 2014 On realizations of polynomial algebras with three generators via deformed oscillator algebras \textit{J. Phys. A: Math. and Theor.} {\bf 47}, 205203.

\bibitem{FH3} Hoque M F, Marquette I and Zhang Y-Z 2015 Quadratic algebra structure and spectrum of a new superintegrable system in $N$-dimension \textit{J. Phys. A : Math. Theor.} {\bf 48}, 185201. 
  
\bibitem{FH4} Hoque M F, Marquette I and Zhang Y-Z 2015 A new family of $N$-dimensional superintegrable double singular oscillators and quadratic algebra $Q(3)\oplus so(n)\oplus so(N-n)$ \textit{J. Phys. A : Math. Theor.} (in press).

\bibitem{Jau1} Jauch J M and Hill E L 1940 On the problem of degeneracy in quantum mechanics \textit{Phys. Rev.} {\bf 57}, 641.

\bibitem{Boy1} Boyer C P and Miller J W 1974 A classification of second-order raising operators for Hamiltonians in two variables \textit{J. Math. Phys.} {\bf 15}, 9.

\bibitem{Eva1} Evans N W and Verrier P E 2008 Superintegrability of the caged anisotropic oscillator \textit{J. Math. Phys.} {\bf 49}, 092902.

\bibitem{Mar1} Marquette I 2010 Superintegrability and higher order polynomial algebras II \textit{J. Phys. A: Math. Gen.} {\bf 43}, 135203.

\bibitem{Kre1} Krein M G 1957 On a continual analogue of a Christoffel formula from the theory of orthogonal polynomials \textit{Dokl. Akad. Nauk SSSR} {\bf 113}, 970.

\bibitem{Adl1} Adler V E 1994 A modification of Crum's method \textit{Theor. Math. Phys.} {\bf 101}, 1381.

\bibitem{Jun1} Junker G 1995 \textit{Supersymmetric Methods in Quantum and Statistical Physics, Springer, New York}.

\bibitem{Dem1} Demircioglu B, Kuru S, Onder M and Vercin A 2002 Two families of superintegrable and isospectral potentials in two dimensions \textit{J. Math. Phys.} {\bf 43}, 2133.

\bibitem{Mar2} Marquette I 2009 Supersymmetry as a method of obtaining new superintegrable systems with higher order integrals of motion \textit{J. Math. Phys.} {\bf 50}, 122102.

\bibitem{Rag1} Ragnisco O and Riglioni D 2010 A family of exactly solvable radial quantum systems on space of non-constant curvature with accidental degeneracy in the spectrum \textit{SIGMA} {\bf 6}, 097.

\bibitem{Mar4} Marquette I 2011 An infinite family of superintegrable systems from higher order ladder operators and supersymmetry \textit{J. Phys.: Conf. Ser.} {\bf 284}, 012047.

\bibitem{Kal2}  Kalnins E G, Kress J M and Miller W Jr 2011 A recurrence relation approach to higher order quantum superintegrability \textit{SIGMA} {\bf 7}, 031.

\bibitem{Cal1} Calzada J A, Kuru S and Negro J 2014 Superintegrable Lissajous systems on the sphere \textit{Eur. Phys. J. Plus} {\bf 129}, 164.

\bibitem{Cal2} Calzada J A, Kuru S and Negro J 2014 Polynomial symmetries of spherical Lissajous systems \textit{e-print arXiv:} {\bf 1404.7066}.

\bibitem{FH1}  Hoque M F, Marquette I and Zhang Y-Z 2016 Recurrence approach and higher rank cubic algebras for the $N$-dimensional superintegrable systems \textit{J. Phys. A: Math. Theor.} {\bf 49}, 125201.

\bibitem{FH2}  Hoque M F, Marquette I and Zhang Y-Z 2016 Recurrence approach and higher rank polynomial algebras for the superintegrable monopole systems \textit{arXiv} {\bf 1605.06213}.

\bibitem{And1} Andrews G E, Askey R and Roy R 1999 Special Functions  \textit{Encyclopedia of Mathematics and its Applications} {\bf 70}, (Cambridge University Press).


\end{thebibliography}
\end{document}